# Enhanced GeSn Microdisk Lasers Directly Released on Si

*Youngmin Kim, Simone Assali, Daniel Burt, Yongduck Jung, Hyo-Jun Joo, Melvina Chen, Zoran Ikonic, Oussama Moutanabbir\*, and Donguk Nam\**

YK and SA contributed equally to this work.

Y. Kim, Dr. D. Burt, Y. Jung, H. J. Joo, M. Chen, Prof. D. Nam
School of Electrical and Electronic Engineering, Nanyang Technological University, 50 Nanyang Avenue, 639798, Singapore
E-mail: dnam@ntu.edu.sg

Dr. S. Assali, Prof. O. Moutanabbir
Department of Engineering Physics, École Polytechnique de Montréal, C.P. 6079, Succ. Centre-Ville, Montréal, Québec H3C 3A7, Canada
E-mail: oussama.moutanabbir@polymtl.ca

Dr. Z. Ikonic
School of Electronic and Electrical Engineering, University of Leeds, Leeds LS2 9JT, UK

Abstract: GeSn alloys are promising candidates for complementary metal-oxide-semiconductor (CMOS)-compatible, tunable lasers. Relaxation of residual compressive strain in epitaxial GeSn has recently shown promise in improving the lasing performance. However, the suspended device configuration that has thus far been introduced to relax the strain is destined to limit heat dissipation, thus hindering the device performance. Herein, we demonstrate that strain-free GeSn microdisk laser devices fully released on Si outperform the canonical suspended devices. This approach allows to simultaneously relax the limiting compressive strain while offering excellent thermal conduction. Optical simulations confirm that, despite a relatively small refractive index contrast between GeSn and Si, optical confinement in strain-free GeSn optical cavities on Si is superior to that in conventional strain-free GeSn cavities suspended in the air. Moreover, thermal simulations indicate a negligible temperature increase in our device. Conversely, the temperature in the suspended devices increases substantially reaching, for instance, 120 K at a base temperature of 75 K under the employed optical pumping conditions. Such improvements enable increasing the operation temperature by ~40 K and reducing the lasing threshold by 30%. This approach lays the groundwork to implement new designs in the quest for room temperature GeSn lasers on Si.

# 1. Introduction

Monolithic photonic-integrated circuits (PICs) have the potential to enable a wide variety of emerging technologies such as light detection and ranging (LiDAR), biochemical sensing, and chip-level optical communications[1]. However, the lack of an efficient, monolithic laser that is compatible with CMOS processing limits the practical realization of these long-sought-after PICs[2]. Germanium (Ge) has been extensively explored for such a laser owing to its CMOS compatibility and near-direct bandgap configuration[3,4]. Among various approaches to achieve the bandgap directness, strain engineering[5–13] and tin (Sn) alloying[14–17] have been considered as the two most promising paradigms.

While lasing action has been observed in strain-engineered Ge at low operating temperatures (<100 K)[18–20], the Sn alloying approach has made significant, steady progress towards achieving lasing at practically high temperature over the past few years[21–27]. Since the first lasing demonstration in GeSn at 90 K[21], much effort has been focused on increasing the operating temperature[21–27]. A major route to this end has been to increase the Sn content to further increase the directness of GeSn alloys[17], which enabled higher operating temperatures reaching 270 K[27]. However, the lasing thresholds at these elevated temperatures are very high (>800 kW cm$^{-2}$ at 270 K)[27]. The exact causes for such high threshold in direct bandgap GeSn lasers have been attributed to the material quality[17]. For instance, it has been suggested that a large content of Sn increases the non-radiative recombination rate, thus leading to the reduction of internal quantum efficiency which influences the lasing threshold significantly[17]. In addition, the Sn alloying is typically accompanied by the compressive strain in the GeSn layer due to the large lattice mismatch between GeSn and Ge buffer layers[15]. Such compressive strain reduces the directness of GeSn[28], thereby hindering the lasing performance. Additionally, the increase in Sn content requires a decrease in growth temperature, which is typically associated with a higher concentration of point defects (vacancies and vacancies complexes) that can also impact the laser performance due to carrier trapping[29,30].

Another route to improve the lasing performance is to simultaneously employ both strain engineering and Sn alloying[28,31]. Recently, a few research groups have made significant progress along this direction by relaxing the compressive strain[22,32] and also by inducing mechanical tensile strain in GeSn[33,34]. Despite the improved directness of strain-engineered GeSn over as-grown compressively strained GeSn, the suspended device configuration, which has thus far been necessary for strain-engineered GeSn[22,32–34], inevitably leads to a very poor thermal conduction property. Obviously, mitigating the latter should lay the ground to fully harness the advantages of strain-engineered GeSn.

In this work, we demonstrate an enhanced performance in GeSn microdisk lasers by combining strain engineering and thermal management. To this end, we selectively removed the Ge buffer layer completely to fully release the strain-free microdisk lasers directly on the underlying Si substrate, which has a relatively high thermal conductivity (~1.3 W cm$^{-1}$ K$^{-1}$ at room temperature). We confirmed that, despite a relatively small refractive index contrast between GeSn and Si, the optical quality (Q) factor in our lasers can be as high as ~11000, which is superior to that in suspended laser structures (~10000). The high optical confinement along with the improved thermal conduction in laser structures fully released on Si allowed observing lasing action at increased operating temperatures up to 110 K compared to the suspended laser devices (<75 K). This operating temperature is higher than other reports for relaxed GeSn at similar Sn contents (~7 at%)[32]. Our thermal simulations using the finite-element method (FEM) along with optical gain modeling provided quantitative evidence for the improved operation temperature for the fully released lasers. The threshold density (<60 kW cm$^{-2}$ at 4 K) was also reduced by ~30% compared to the suspended devices.

## 2. Material Characterization

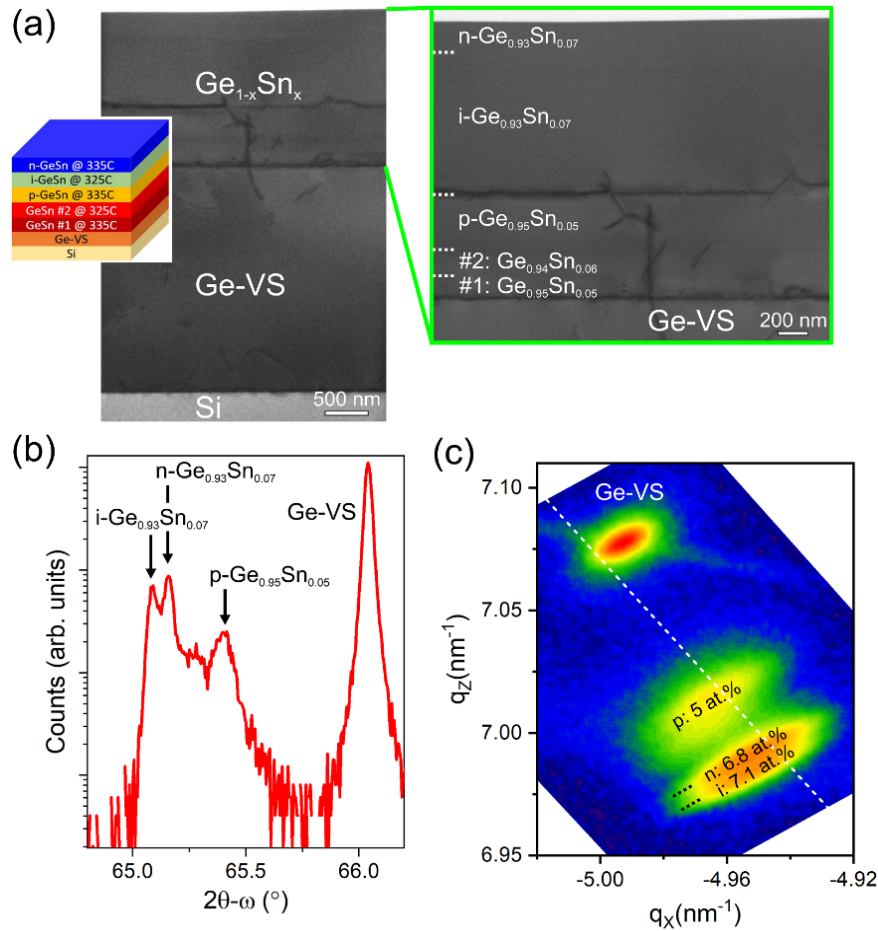

**Figure 1.** (a) Cross-sectional TEM image of GeSn p-i-n/Ge-VS heterostructure grown on a Si wafer. (b) 2θ-ω scan around the (004) X-ray diffraction order. (c) RSM-XRD around the asymmetrical (224) reflection.

The GeSn p-i-n heterostructure was grown on a 4-inch Si (100) wafer in a low-pressure chemical vapor deposition (CVD) reactor using ultra-pure $H_2$ carrier gas. 10 % monogermane ($GeH_4$) and tin-tetrachloride ($SnCl_4$) precursors, while p-type and n-type doping was controlled using 1.6 % Diborane ($B_2H_6$) and 1 % Arsine ($AsH_3$), respectively. A 2.6 µm-thick Ge virtual substrate (VS) was grown by a two-temperature step process at 450 and 600 °C followed by post-growth thermal cyclic annealing above 800 °C. The GeSn layers were grown at a reactor pressure of 50 Torr, constant $H_2$ flow and $GeH_4$ molar fraction ($1.2 \cdot 10^{-2}$), and the composition was controlled by the temperature change[16,35]. In addition, the molar fraction of the $SnCl_4$ precursor ($9 \times 10^{-6}$ at 335 °C) was reduced by ~20 % during each temperature step to compensate for the reduced GeH$_4$ decomposition as temperature decreases. Two GeSn buffer layers with compositions of 5 at% and 6 at% and thicknesses of 130 nm and 200 nm were grown at 335 °C (#1) and 325 °C (#2), respectively. The subsequent GeSn p-i-n layer with a composition of 5-7-5 at% and a thickness of 290-830-330 nm was grown at 335-325-335 °C with Ge/Sn ratio in gas phase of 1290-1610-1290. B$_2$H$_6$ and AsH$_3$ were introduced for the growth of p-GeSn (B/Ge ratio in gas phase 2×10$^{-3}$) and n-GeSn regions (As/Ge ratio in gas phase 9×10$^{-5}$), respectively. The active doping in both p-GeSn and n-GeSn layers is estimated to be higher than 1×10$^{19}$ cm$^{-3}$.

The cross-sectional transmission electron micrograph (TEM) of the GeSn p-i-n heterostructure with a total thickness of 1800 nm grown on Ge-VS/Si is shown in Figure 1a. Defective layers are visible at the Ge$_{0.95}$Sn$_{0.05}$(#1)/Ge and p-Ge$_{0.95}$Sn$_{0.05}$/i-Ge$_{0.93}$Sn$_{0.07}$ interfaces, while high crystalline quality is obtained in the upper region of the material stack. The GeSn multi-layer heterostructure does not display a significant increase in threading dislocation density (TDD) compared to the Ge-VS (TDD >10$^7$ cm$^{-2}$), however gliding of pre-existing threading dislocations in the Ge-VS will occur during the lattice-mismatched GeSn growth[35]. When the thickness of the (#1-2, p) GeSn layers with an average composition of ~5 at% exceeds the critical thickness, high density of misfit dislocations nucleate at the interface with the Ge-VS to reduce the compressive strain in the system. Next, when the lattice-mismatched growth of the (i, n) Ge$_{0.93}$Sn$_{0.07}$ layers exceeds again the critical thickness with respect to the Ge$_{0.95}$Sn$_{0.05}$ substrate, a network of misfit dislocations forms at the interface with the latter to partially relax the strain in the system. The crystalline quality of the GeSn p-i-n heterostructure was further evaluated using X-ray diffraction spectroscopy (XRD). 2θ-ω measurements around the (004) XRD order and Reciprocal Space Mapping (RSM) measurements around the asymmetrical (224) XRD peak were performed as shown in Figure 1b and 1c, respectively. The small (in-plane) tensile strain $\varepsilon_\parallel$ < 0.2 % in the Ge-VS is due to the thermal cyclic annealing process, while the plastic relaxation in the GeSn multi-layer heterostructure results in a very low compressive strain ($\varepsilon_{\parallel} < -0.2$ %).

## 2. Device Fabrication and Simulation

Microdisk optical resonators were patterned using photolithography, followed by anisotropic dry etching using reactive ion etching (RIE) with $Cl_2$ chemistry to etch the GeSn layer and expose the Ge buffer layer. Isotropic dry etching was then employed in an RIE chamber using $CF_4$ chemistry, which selectively undercuts the Ge buffer layer whilst leaving the GeSn layers intact[36]. By controlling the isotropic etching time for each sample, both suspended (Figure 2a) and released (Figure 2b) microdisks were successfully fabricated as shown in the scanning electron microscopy (SEM) images. For the released microdisks, the underlying Ge layer was completely etched by etching for a longer time, thus allowing the microdisks to collapse onto the Si substrate. For this study, optical characterization was performed on a set of two different microdisks with a diameter of 11 μm.

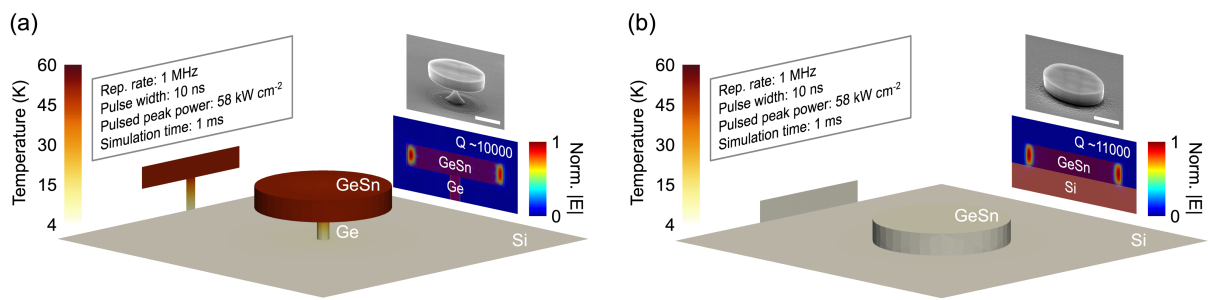

**Figure 2.** FEM thermal simulations, FDTD optical simulations and SEM images of the (a) suspended and (b) released microdisks. Scale bars in the SEM images are 5 μm.

To quantify the difference in thermal dissipation between the two microdisks, we first performed thermal simulations using 3D finite element method (FEM). In the simulations, the center of a microdisk was illuminated with pulsed optical pumping of 58 kW cm$^{-2}$ with a beam spot diameter of 10 μm to mimic the experimental conditions. The pulse duration and repetition rate were set to 10 ns and 1 MHz, respectively. The base temperature was set to 4 K. The thermal conductivity of GeSn is set to 0.1 W cm$^{-1}$ K$^{-1}$ that is lower than that of Ge (0.58 W cm$^{-1}$ K$^{-1}$) due to alloy-induced phonon scattering effects[37]. The equilibrium temperature profile of the suspended microdisk shows a significant temperature increase from 4 K to 60 K in the GeSn layers (Figure 2a), whereas the released microdisk shows negligible heating (Figure 2b), thus confirming the excellent thermal dissipation through the Si substrate.

Finite-difference time-domain (FDTD) simulations were conducted to compare the optical confinement in the two microdisks as also shown in Figure 2a and 2b, respectively. The normalized electric field intensity is overlapped with the structure to show the spatial distribution of the optical mode. Despite a relatively small refractive index contrast between GeSn and Si, the Q-factor of the released microdisk (~11000) were found to be comparable to that of the suspended microdisk (~10000). The slightly lower Q-factor in the suspended microdisk can be attributed to the optical field leaking into the Ge pillar, which was evidenced in detailed optical mode simulations (see Figure S1 in Supporting Information). These comprehensive thermal and optical simulations show the potential of the released GeSn microdisks to improve lasing performance.

## 3. Optical Characterization

PL studies were conducted to investigate and compare the lasing characteristics of the two sets of microdisks. The samples were loaded into a closed-cycle helium cryostat operating between 4 K and 300 K to allow for temperature-dependent studies. A 1550-nm wavelength pulsed laser was focused onto the sample using a ×15 reflective objective lens, which resulted in a beam spot size diameter of ~10 μm. The repetition rate and pulse width were set to 1 MHz and 10 ns, respectively. The emission was collected by the same reflective objective lens and coupled into a Fourier transform infrared (FTIR) spectrometer with an extended InGaAs detector with a cut-off wavelength of 2.4 μm.

Figure 3a shows the emission spectra of both the suspended and released microdisks with increasing pump power densities at a temperature of 4 K. For both microdisks, only broad spontaneous emission can be observed at low pump power densities (black curves in Figure 3a and 3b). As the pump power density increases to ~60 kW cm$^{-2}$, a single sharp lasing peak arises for the released microdisk at ~2215 nm (blue curve in Figure 3b) while the suspended microdisk only shows two weak optical resonances (blue curve in Figure 3a). At higher pump power densities, both devices show clear lasing behavior. For the suspended microdisk, two distinct lasing modes appear at ~2217 nm and ~2226 nm. The lasing mode at ~2217 nm disappears due to mode competition, resulting in a single lasing mode at 2226 nm as the pump power density is further increased (red curve in Figure 3a). In contrast, the released microdisk show a second smaller lasing mode at ~2205 nm for a higher pump power density (red curve in Figure 3b) possibly due to gain broadening originating from band-filling effects[33].

Further evidence of lasing is observed in the light-in-light-out (L-L) curves of the suspended (blue) and released (red) microdisks in Figure 3c. Clear threshold behavior can be observed both in the linear and double-logarithmic (inset) plots. An S-shaped curve in the double-logarithmic plot represents a clear signature of lasing. The threshold power densities were extracted as 76.9 kW cm$^{-2}$ and 58.4 kW cm$^{-2}$ for the suspended and released microdisks, respectively. The lower threshold power density for the latter can be attributed mainly to improved thermal management since other factors such as the starting material and degree of strain relaxation were the same for the two devices. As also evidenced in the thermal simulations (Figure 2), compared to the suspended microdisk, the released microdisk can be heated significantly less under optical pumping, resulting in lower material losses and the onset of a positive net gain at a lower pump power density. A clear reduction in the full width at half-maximum (FWHM) can also be observed for both the suspended and released microdisks, as shown in Figure 3d. The FWHM decreased from a value of ~2.0 nm (400 µeV) to ~0.7 nm (180 µeV) for both microdisks.

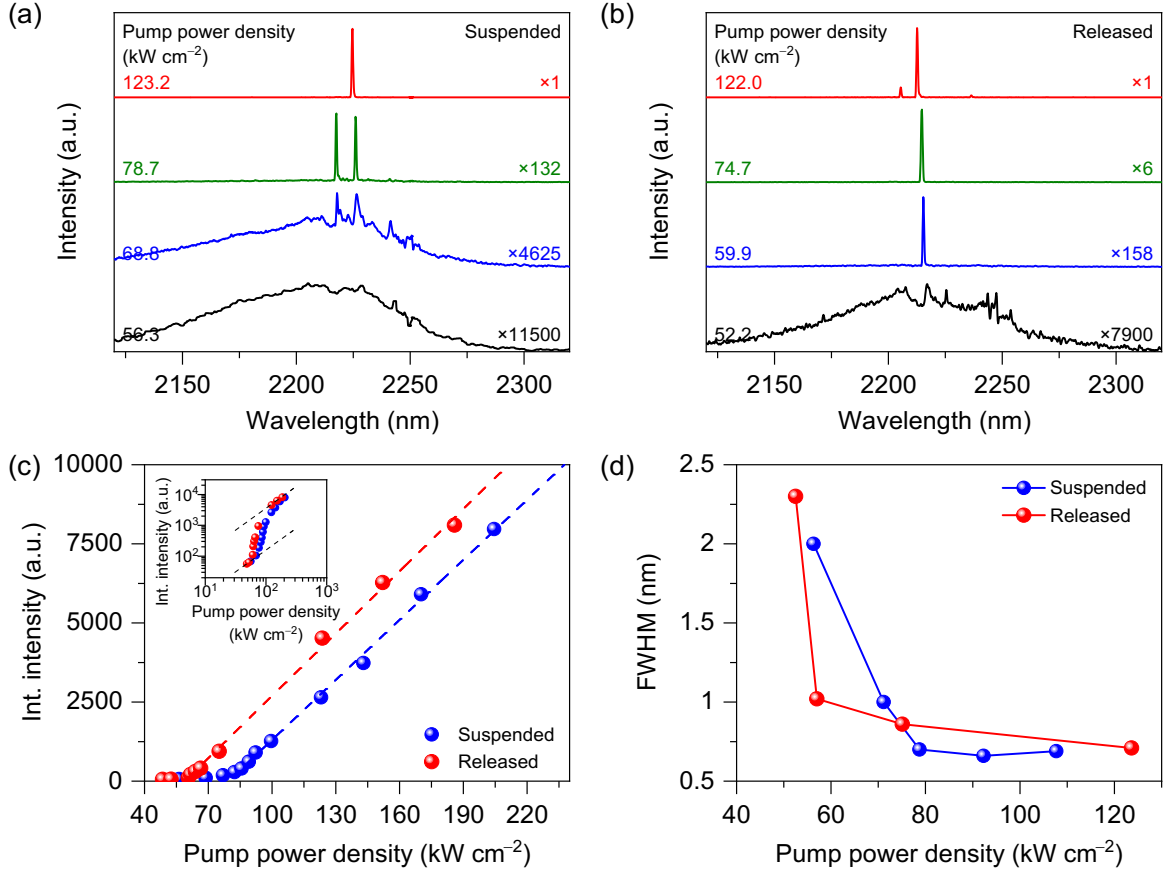

**Figure 3.** Emission spectra of the (a) suspended and (b) released microdisks at 4 K. (c) L-L curves for the suspended and released microdisks at 4 K. The inset shows the double-logarithmic plot. (d) FWHM as a function of pump power density for the suspended and released microdisks at 4 K.

Figure 4a presents the normalized emission spectra of the suspended (blue) and released (red) microdisks at 75 K for a fixed pump power density of 74.2 kW cm$^{-2}$. In contrast to the released microdisk showing a clear lasing peak, the suspended microdisk only shows a broad spontaneous emission. The L-L curves in the inset to Figure 4a exhibits the contrasting behaviors of the two devices. The FEM thermal simulations with the same pump power density of 74.2 kW cm$^{-2}$ for the two devices show that the suspended microdisk heats up to ~130 K, whereas the temperature of the released microdisk remains largely unchanged at ~75 K (Figure 4b). Thus, the absence of lasing in the suspended microdisk can be ascribed to increased material losses owing to the increased temperature. The thermal analyses along with experimental evidence in Figure 4 clearly display the superior thermal management in released microdisk lasers.

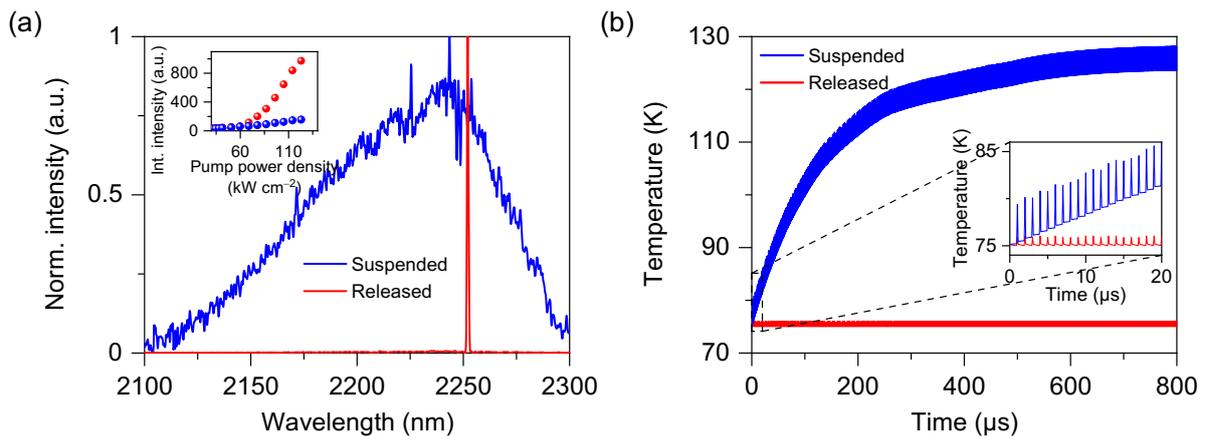

**Figure 4.** (a) Normalized emission spectra and L-L curves (inset) of the suspended and released microdisks at 75 K. (b) Simulated temperature as a function of time for the suspended and released microdisks at 75 K. Inset: zoomed-in version of Fig. 3(b) for the initial, short time range, which manifests the device temperature variation due to the pulsed pumping.

The lasing emission in the released microdisk persisted at elevated temperatures of up to 130 K as shown in Figure 5a. The spectra in Figure 5a were acquired at a fixed pump power density of 123.1 kW cm$^{-2}$. In contrast to clear single-mode lasing at 75 K (red) and 110 K (green), the spectrum at 130 K (blue) shows two smaller peaks. At 150 K (black), only broad spontaneous emission was observed. Figure 5b shows the L-L curves of the released microdisk at various temperatures. Lasing action at 75 K and 110 K can be evidenced by clear threshold behaviors. At 130 K, the integrated PL intensity shows a superlinear increase at ~150 kW cm$^{-2}$, but rolls-over before reaching the clear lasing regime possibly due to the elevated device temperature at high excitation levels (>300 kW cm$^{-2}$).

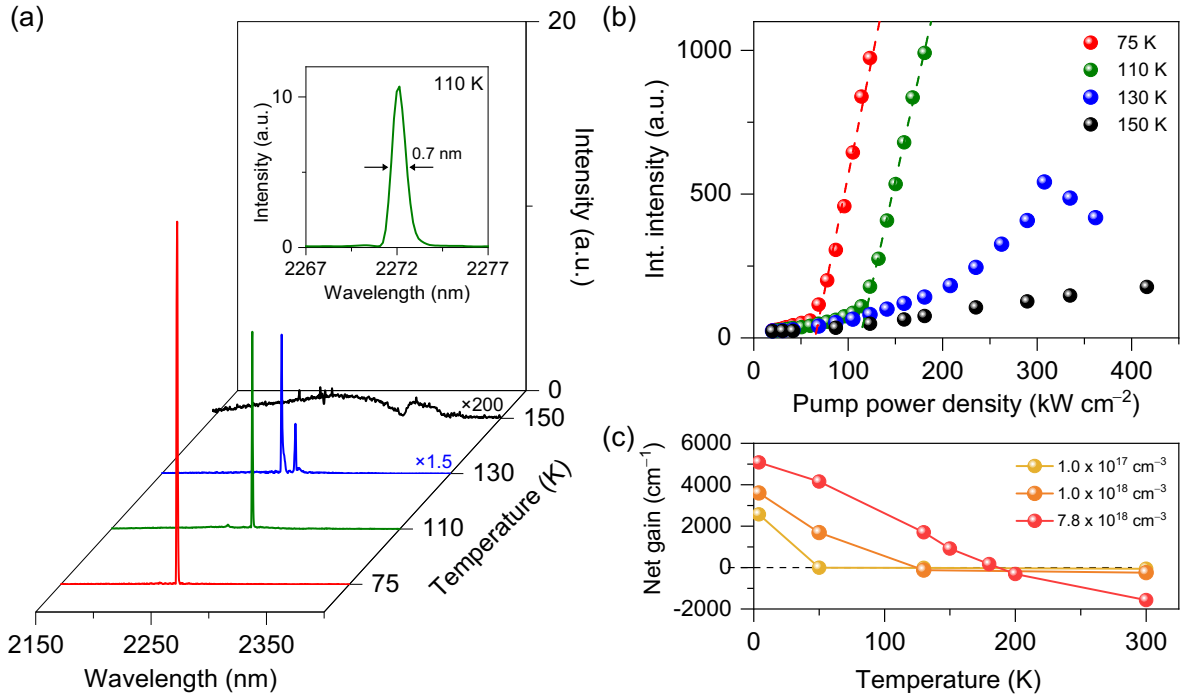

**Figure 5.** (a) Emission spectra of the released microdisk at various temperatures. Inset shows the zoomed-in lasing peak at 110 K. (b) L-L curves for the released microdisk at various temperatures. (c) Calculated net gain for different temperatures at various injected carrier densities of $1 \times 10^{17}$ cm$^{-3}$ (yellow), $1 \times 10^{18}$ cm$^{-3}$ (orange) and $1 \times 10^{19}$ cm$^{-3}$ (red).

The effect of the device temperature on the lasing action was further studied via theoretical modeling (Figure 5c). We used the k·p method to model the bandstructure of GeSn, and then calculated the optical interband gain[38], and also accounted for the free carrier absorption[39] to find the net gain. Figure 5c shows optical net gains for different temperatures with various injected carrier densities of $1.0 \times 10^{17}$ cm$^{-3}$, $1.0 \times 10^{18}$ cm$^{-3}$ and $1.0 \times 10^{19}$ cm$^{-3}$. It is worth noting that the pump power range used to observe the superlinear growth in the integrated PL intensity at 130 K (blue in Figure 5b) is between 100 kW cm$^{-2}$ and 300 kW cm$^{-2}$, which correspond to injection carrier densities of $2.6 \times 10^{18}$ cm$^{-3}$ and $7.8 \times 10^{18}$ cm$^{-3}$, respectively. As shown in Figure 5c, for the injected carrier density between $1.0 \times 10^{18}$ cm$^{-3}$ and $1.0 \times 10^{19}$ cm$^{-3}$, net gain can be achieved in the temperature range between 130 K and 210 K. The simulated results display that the net gain increases at higher pump powers (from yellow to red) but decreases when the device temperature is increased. Additional thermal simulations showed that the temperature of the released devices also rises from ~130 K to ~160 K upon the high pump excitation of 300 kW cm$^{-2}$ (see Figure S2 in Supporting Information). At an injected carrier density of $7.8 \times 10^{18}$ cm$^{-3}$ that corresponds to a pump power of 300 kW cm$^{-2}$, the net gain quickly decreases as the simulated temperature is increased. Although the net gain remains positive up to ~180 K, lasing should stop at a temperature lower than 180 K because the net gain should exceed the optical losses of the microdisk resonators. Our experimental results along with theoretical analyses provide insight towards increasing the operating temperature of GeSn lasers by improving the thermal dissipation. For example, deposition of a thermally conductive and optically insulating layer (e.g., $Al_2O_3$) can be explored in a future study to further improve the thermal management in GeSn microdisk lasers.

## 4. Conclusion

In conclusion, we have experimentally demonstrated that strain-free GeSn microdisk lasers fully released on Si can achieve clear lasing action with an operation temperature reaching 110 K, which is higher than other reports for relaxed GeSn at a similar Sn content (~7 at%). The observed threshold density in the fully released laser devices (<60 kW cm$^{-2}$ at 4 K) was lower than that in the suspended devices (~80 kW cm$^{-2}$ at 4 K). The comprehensive thermal and optical simulations indicate that these substantial improvements can be ascribed to the excellent thermal dissipation and optical confinement in released microdisks. The k·p calculations also highlighted the detrimental effect of the device temperature on the net gain, which can be mitigated in the released microdisks. This work highlights the importance of achieving efficient heat dissipation to improve the lasing performance in GeSn microdisk lasers. These insights and associated systematic studies provide additional degrees of freedoms to control the basic properties of GeSn device structures and implement new designs to increase the laser operating temperature towards the ultimate objective of realizing room-temperature CMOS-compatible monolithic lasers.

**Supporting Information**

Supporting Information is available from the Wiley Online Library.

**Acknowledgements**

The work carried out in Montréal was supported by NSERC Canada (Discovery, SPG, and CRD Grants), Canada Research Chairs, Canada Foundation for Innovation, Mitacs, PRIMA Québec, and Defence Canada (Innovation for Defence Excellence and Security, IDEaS). The research performed at NTU was supported by Ministry of Education, Singapore, under grant AcRF TIER 1 2019-T1-002-050 (RG 148/19 (S)). The research of the project was also supported by Ministry of Education, Singapore, under grant AcRF TIER 2 (MOE2018-T2-2-011 (S)). This work is also supported by National Research Foundation of Singapore through the Competitive Research Program (NRF-CRP19-2017-01). This work is also supported by National Research Foundation of Singapore through the NRF-ANR Joint Grant (NRF2018-NRF-ANR009 TIGER). This work is also supported by the iGrant of Singapore A*STAR AME IRG (A2083c0053).The authors would like to acknowledge and thank the Nanyang NanoFabrication Centre (N2FC).

# Supporting Information

**Enhanced GeSn Microdisk Lasers Directly Released on Si**

*Youngmin Kim, Simone Assali, Daniel Burt, Yongduck Jung, Hyo-Jun Joo, Melvina Chen, Zoran Ikonic, Oussama Moutanabbir*, and Donguk Nam**

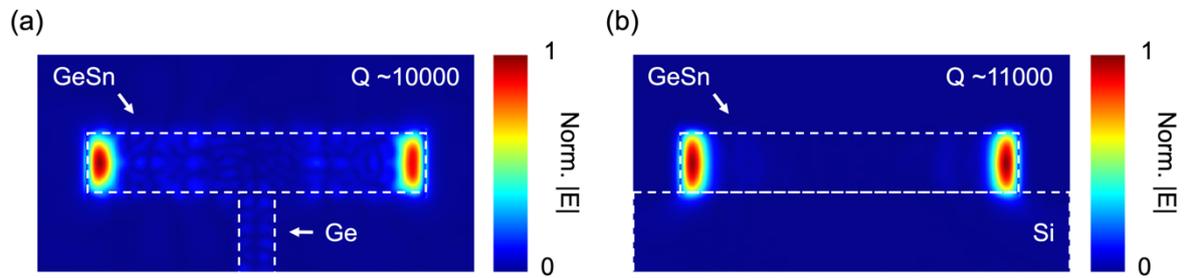

Figure S1. FDTD optical simulation results for (a) suspended and (b) released microdisk laser devices.

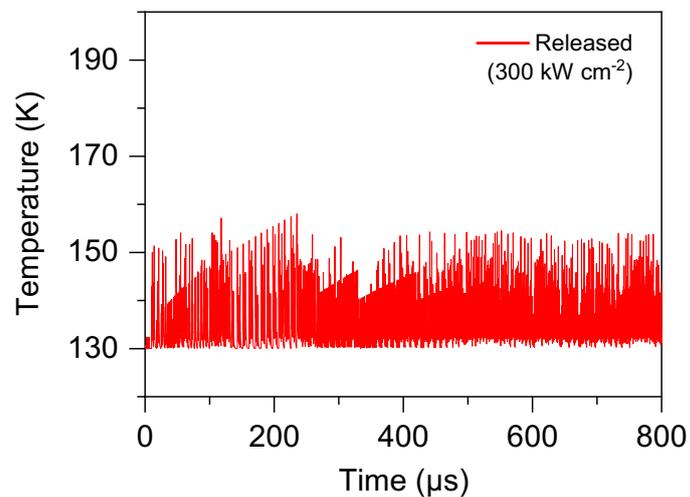

Figure S2. Simulated temperature as a function of time for the released microdisk at a base temperature of 130 K and at the high pump excitation of 300 kW cm$^{-2}$.

16